\documentclass[%
 preprint,
superscriptaddress,
nofootinbib,
bibnotes,
 amsmath,amssymb,
 aps,
prl,
longbibliography,
]{revtex4-1}

\usepackage{graphicx}
\usepackage{dcolumn}
\usepackage{bm}

\usepackage{gensymb}
\usepackage[colorlinks=true,linkcolor=blue,citecolor=blue]{hyperref}

\usepackage{orcidlink}

\newcommand{\orcidJuan}{0000-0003-3756-5016}
\newcommand{\orcidMary}{0000-0003-0320-5768}

\begin{document}


\title{Spectral thermodynamics of a soliton heat engine}

\author{M. Ahumada\orcidlink{\orcidMary}}
\affiliation{Departamento de F\'isica, Universidad de Santiago de Chile,
Av. Victor Jara 3493, Estaci\'on Central, Santiago, Chile}

\author{J. F. Mar\'in\orcidlink{\orcidJuan}}
\email[]{j.marinm@utem.cl}
\affiliation{Departamento de Física, Facultad de Ciencias Naturales, Matemática y del Medio Ambiente, Universidad Tecnológica Metropolitana, Las Palmeras 3360, Ñuñoa 780-0003, Santiago, Chile.}

\begin{abstract}
We demonstrate a thermodynamic engine whose working substance is a sine-Gordon soliton in a heterogeneous current-driven Josephson junction. We show that solitons can act as thermodynamic working substances whose internal spectral structure enables energy conversion beyond conventional few-level engines. By dynamically deforming the soliton using a controllable dipole current, the internal bound-state spectrum of the soliton can be engineered in time, enabling a finite-time Carnot-like cycle based on spectral control, in close analogy with quantum heat engines. Mapping the instantaneous nonlinear field configuration to an effective Schr\"odinger operator, we reveal how bound states appear, approach the continuum threshold, and disappear during the cycle. Comparing three thermodynamic descriptions (full nonlinear field dynamics, a coarse-grained mesoscopic model, and a two-level spectral model), we show that few-level descriptions systematically underestimate the engine performance. The enhanced efficiency arises from the extended nature of the soliton, whose internal spectral degrees of freedom provide additional energy storage and transfer channels. Our results reveal a general thermodynamic principle: extended nonlinear excitations with particle-like behavior can serve as tunable working media, whose internal spectral degrees of freedom provide additional reversible channels for energy storage and transfer beyond those of few-level systems. 
\end{abstract}

\maketitle


Heat engines convert thermal energy into work through cyclic processes and constitute a central paradigm of thermodynamics \cite{Huang1965, Kardar2007}. Advances in quantum thermodynamics have explored heat engines whose working medium consists of a small number of quantum levels, such as two- or three-level systems, harmonic oscillators, or qubits \cite{Scovil1959, Geusic1967, Alicki1979, Kosloff1984, Kosloff2014, Peterson2019}. These quantum heat engines reveal how coherence, correlations, and discrete spectra can enhance thermodynamic performance \cite{Klatzow2019, Barrios2017, Ahumada2023}. Most theoretical and experimental realizations of microscopic heat engines, however, share a common assumption: the working medium possesses only a few degrees of freedom, typically a small number of discrete energy levels \cite{Quan2007, Esposito2010, Ma2024, Ono2020, Cangemi2024}. While such models capture essential thermodynamic principles, they overlook a broad class of physical systems that lie between microscopic quantum devices and macroscopic classical media, including devices exhibiting macroscopic quantum effects \cite{Devoret1984, Martinis1985, Devoret1985}. Many physical systems capable of storing and redistributing energy are complex or spatially extended media, such as glasses, disordered solids, and nonlinear fields, in which energy flows through a hierarchy of collective excitations and emergent modes \cite{Brazhkin2014, Scalliet2019, Gonzalez2002}. Indeed, energy can be stored and transferred by extended nonlinear excitations (such as vortices, domain walls, and solitons) that behave as particle-like entities while possessing internal structure \cite{Gonzalez1992, Gonzalez2003, Gonzalez2006, Ustinov2002}.

In this work, we show that solitons can act as thermodynamic working substances and can outperform conventional few-level engines. We consider a fluxon trapped in a long Josephson junction (JJ) driven by an externally controlled current dipole, which dynamically deforms the soliton profile. This deformation reshapes the effective potential experienced by fluctuations around the soliton, thereby engineering the bound-state spectrum of its internal modes in time. Using this spectral control, we construct a Carnot-like cycle in which the appearance and disappearance of internal bound states govern heat exchange and work extraction. By comparing three complementary thermodynamic descriptions of the engine: full nonlinear field dynamics, a mesoscopic coarse-grained model, and a reduced spectral model, we show that few-level descriptions systematically underestimate the performance of the soliton engine. The enhanced efficiency arises from the extended and structured nature of the topological soliton, whose particle-like behavior and internal spectral structure provide additional channels for storing and redistributing energy beyond those captured by simple few-level models.



\begin{figure*}
\includegraphics[width=1\textwidth]{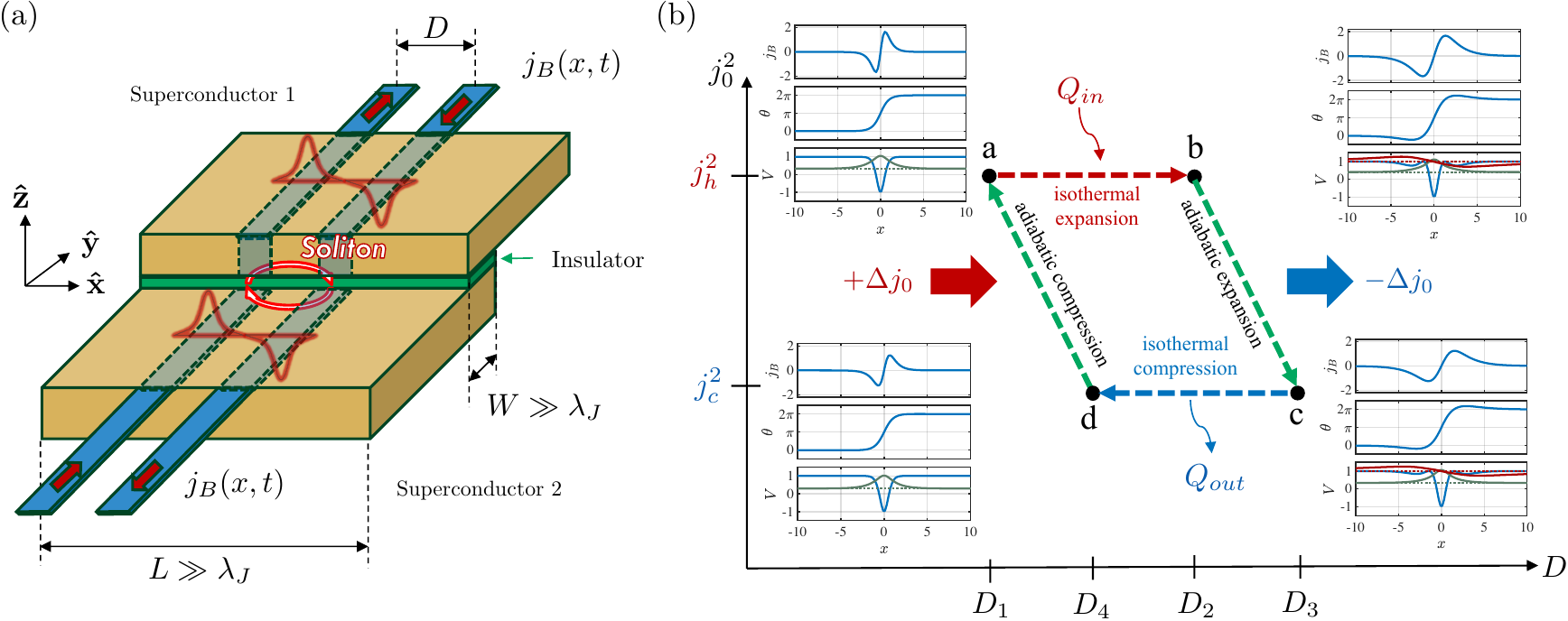}
\caption{A soliton trapped in a Josephson junction as a controllable thermodynamic working substance. (a) Device schematic. A long ($L\gg\lambda_J$) and quasi one-dimensional ($\lambda_J\ll W\ll L$) junction with a soliton trapped by a dipole current with dipole distance $D$ and a heterogeneous density current profile $j_B(x,t)$. (b) Carnot cycle in the control parameter space $(D, j_o^2)$ \cite{SupMat}. The insets show the dipole current, the soliton profile, and the effective potential well supporting bound states at each stage. Numerically obtained eigenfunctions and pseudo-energy levels are shown in solid and dashed lines, respectively, for the ground (green) and excited (red) states. }
\label{Fig:01} 
\end{figure*}

We consider a long quasi one-dimensional JJ of length $L$ and width $W$, as shown in Fig.~\ref{Fig:01}(a), whose dynamics are described by the driven-damped sine-Gordon (sG) equation~\cite{Ustinov1998, Malomed2004, CuevasMaraver2014}
\begin{equation}
    \label{Eq:01}
    \partial_{tt}\theta-\partial_{xx}\theta+\gamma\partial_t\theta+\sin\theta=j_B(x,t)\,,
\end{equation}
where space $x$ and time $t$ are measured in units of the Josephson length $\lambda_J$ and plasma frequency, respectively, $\theta$ is the superconducting phase difference across the junction, $\gamma=0.01$ is the damping coefficient, and $j_B$ represents an externally applied bias current. Fluxons in the junction correspond to topological soliton solutions of the sG equation and carry one quantum of magnetic flux \cite{Barone1982, Peyrard2004}. To trap and manipulate a fluxon, we introduce a dipole current profile
\begin{equation}
   \label{Eq:02} 
   j_B(x,t)=j_0(t)\tanh[B(t)\,x]\,\mbox{sech}[B(t)\,x]\,,
\end{equation}
which produces a localized potential capable of pinning a soliton $\theta_f(x,t)=4\arctan\exp[B(t)\,x]$, where $B(t)$ is a time-dependent shape parameter inversely proportional to the soliton width. The parameters $B(t)$ and $j_0(t)$ control the effective dipole distance and current intensity, thereby dynamically deforming the soliton profile \cite{CastroMontes2020, CastroMontes2025}. Small perturbations around the instantaneous soliton configuration can be described by the effective Schr\"odinger operator
\begin{equation}
\label{Eq:03}
    \hat{H}(t)=-\partial_{xx}+\cos\theta(x,t)\,.
\end{equation}

The potential $V(x,t)=\cos\theta(x,t)$ forms a well whose bound states correspond to internal modes of the soliton \cite{Gonzalez2002}. In the stationary limit, this potential reduces to the analytically solvable P\"oschl-Teller form \cite{Gonzalez2003}, allowing controlled engineering of the bound-state spectrum.

By dynamically varying $B(t)$ and $j_0(t)$, we construct a cyclic protocol that implements the finite-time Carnot-like engine, as shown in Fig.~\ref{Fig:01}(b) and the Supplemental Video \cite{SupMat}. The cycle is defined in the plane spanned by the dipole distance $D(t)=\mbox{arctanh}(2\sqrt{2}/3)/B(t)$ and the squared current intensity $j_0^2(t)$, analogous to the volume-temperature plane of a conventional Carnot cycle. During the cycle, the soliton potential well undergoes controlled deformations that smoothly modify the bound-state spectrum. The driving protocol is implemented via sigmoidal deformations in time of the form $\sigma(t; t_0)=1/(1+\exp[(t-t_0)/\tau]),
$ with characteristic time $\tau=20$ in each of the four equally timed strokes, yielding a total cycle duration $T=4000$. Four representative configurations are shown in the insets of Fig.~\ref{Fig:01}(b). During the first stroke, a $\to$ b, $B(t)$ changes sigmoidally from $\sqrt{8/3}$ to $3/\sqrt{20}$ and $j_o=10/3=$const., the potential widens, and a new bound state enters the well, corresponding to the excitation of an internal soliton mode. In the subsequent adiabatic expansion b $\to$ c, both $B(t)$ and $j_0(t)$ change sigmoidally from  $3/\sqrt{20}$ to $1/\sqrt{3}$ and from $10/3$ to $5/2$, respectively, and the excited level approaches the continuum threshold. During the compression stroke c $\to$ d, there is a sigmoidal evolution of $B(t)$ from $1/\sqrt{3}$ to $3/\sqrt{5}$ with $j_o=5/2=$const., and the excited state merges into the continuum, releasing its energy as radiation. Finally, the potential returns to its original shape, completing the cycle with the stroke d $\to$ a through a sigmoidal evolution of $B(t)$ and $j_0(t)$ from $3/\sqrt{5}$ to $\sqrt{8/3}$ and from $5/2$ to $10/3$, respectively. Details on the implemented protocols of the Carnot cycle are provided in a Companion Article \cite{CompanionArticle}.

\begin{figure}
    \centering
    \includegraphics[width=\linewidth]{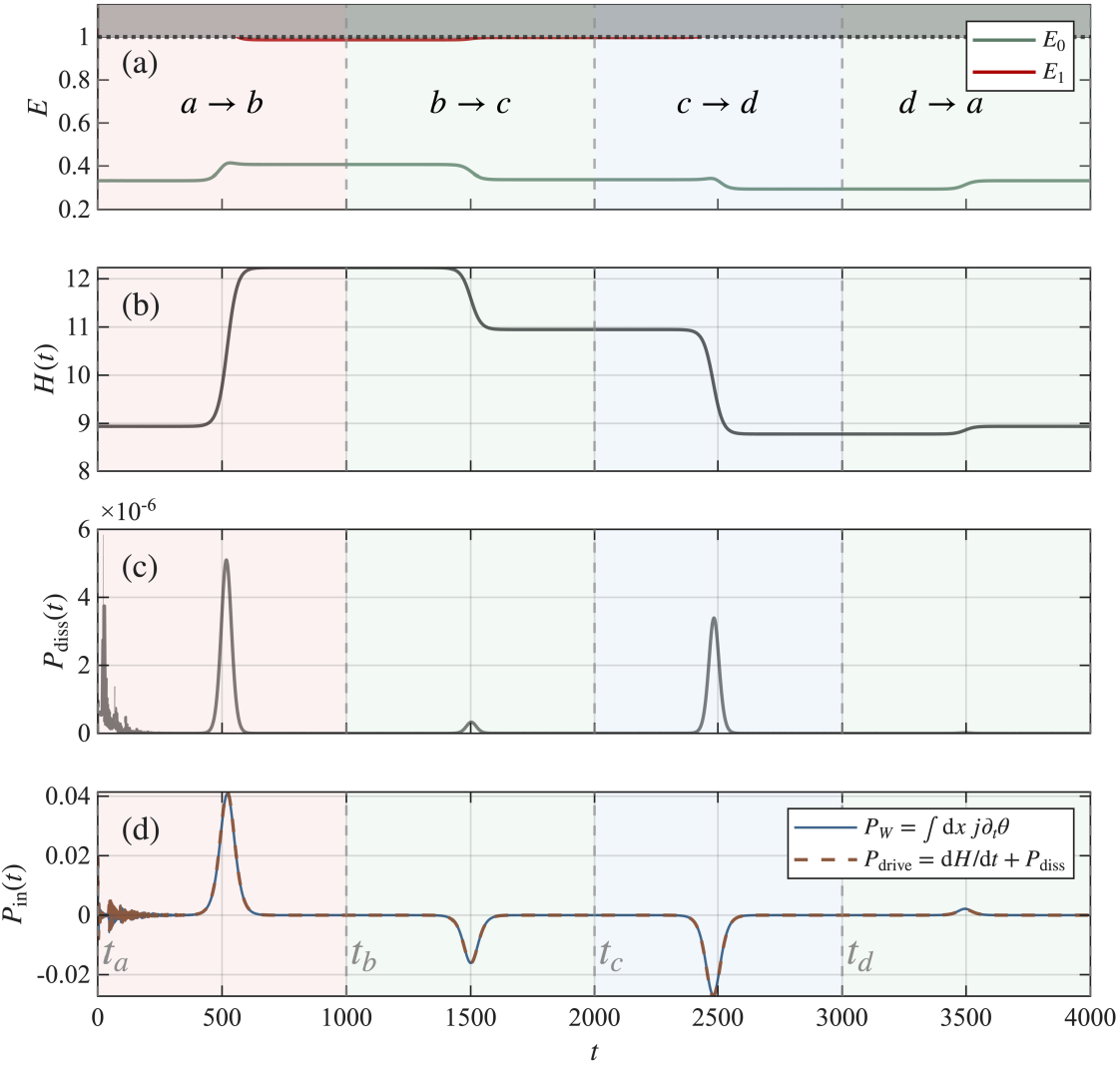}
    \caption{Time resolved energetics of the engine. The four strokes are delineated by vertical dashed lines. (a) Instantaneous spectrum of the soliton, (b) field energy, (c) dissipated power, and (d) work input power using two formulae: the integral of the dipole drive times the field velocity and $P_{\text{drive}}$.}
    \label{Fig:02}
\end{figure}

The instantaneous bound-state energies obtained from the sG field simulations are shown in Fig.~\ref{Fig:02}(a).  The spectral evolution reveals how internal modes appear, approach the continuum threshold (gray region in Fig.~\ref{Fig:02}(a)), and disappear during the thermodynamic cycle. These spectral transitions provide the microscopic mechanism for heat exchange between the soliton and the surrounding field modes. The continuum radiation modes and dissipative environment act as effective thermal reservoirs that absorb or supply energy during the spectral transitions, thereby playing the role of cold and hot baths in the cycle. 

To quantify the performance of the soliton engine, we evaluate its efficiency using three complementary descriptions. First, we compute the efficiency directly from the full nonlinear field dynamics using the energy balance of the sG model. The field energy
\begin{equation}
    \label{Eq:04}
    H(t)=\int\mbox{d}x\,\left\{\frac{1}{2}\left[\left(\partial_t\theta\right)^2+\left(\partial_x\theta\right)^2\right]+1-\cos\theta\right\},
\end{equation}
together with the dissipated power $P_{\text{diss}}(t)=(1/2)\int\mbox{d}x\,\gamma(\partial_t\theta)^2$, determines the heat exchanged during the cycle. The power injected by the dipole current follows from the energy balance relation $P_{\text{drive}}(t)=\mbox{d}H/\mbox{d}t+P_{\text{diss}}(t)$. Figure~\ref{Fig:02}(b-d) summarizes the time-resolved energetics of the engine within this model. Energy, dissipation, and input powers increase markedly during the hot stroke (red region), while heat is strongly dissipated during the cold stroke (blue region). During adiabatic compression (green region $d\rightarrow a$), the field energy increases slightly and then returns to its initial value [Fig.~\ref{Fig:02}(b)]. The system is inherently nonconservative: it operates as a nonequilibrium-driven dissipative system, in which the external drive compensates for small irreversible losses. The dissipation power [Fig.~\ref{Fig:02}(c)] confirms that $P_{\text{diss}}\simeq0$ during the adiabatic strokes (green regions). Dissipation persists throughout the cycle, however, with short peaks between strokes associated with soliton deformation. Finally, Fig.~\ref{Fig:02}(d) shows the input work power $P_{\text{in}}(t)$, with excellent agreement between the two evaluation methods. The efficiency obtained from the full nonlinear field dynamics is, therefore,
\begin{equation}
\label{Eq:05}
\eta_c=1-\frac{Q_{\text{out}}}{Q_{\text{in}}}=1-\frac{\oint\mbox{d}t\,P_{\text{diss}}(t)}{\int_{t_\text{a}}^{t_\text{b}}\mbox{d}t\,P_{\text{drive}}(t)}\,.
\end{equation}

Second, we introduce a spectral two-level description in which the internal energy of the soliton is approximated by the bound-state spectrum of the effective Schr\"odinger operator of Eq.~\eqref{Eq:03}. In this picture, the internal energy of the working medium is expanded as $U(t)=\sum_np_n(t)E_n(t)$ and the populations of the bound states are assumed to follow the Boltzmann statistics, $p_n(t)=\exp[-\beta E_n(t)]/\sum_n\exp[-\beta E_n(t)]$, during the isothermal strokes. We use $\beta=1/k_BT_C$ for the cold bath and $\beta=1/k_B(9T_C)$, with $k_B$ the Boltzmann constant and $T_C=19\,\mbox{mK}$ the typical operational temperature achieved in superconducting circuits \cite{Ahumada2023}. Work is associated with changes in the spectrum for fixed populations, whereas heat corresponds to changes in populations for fixed spectrum \cite{Bender2000}. The efficiency is then $\eta_{\text{TLS}}=W/Q_{\text{in}}$, with work and heat obtained from the spectral decomposition of $U(t)$.

\begin{figure}
    \centering
    \includegraphics[width=\linewidth]{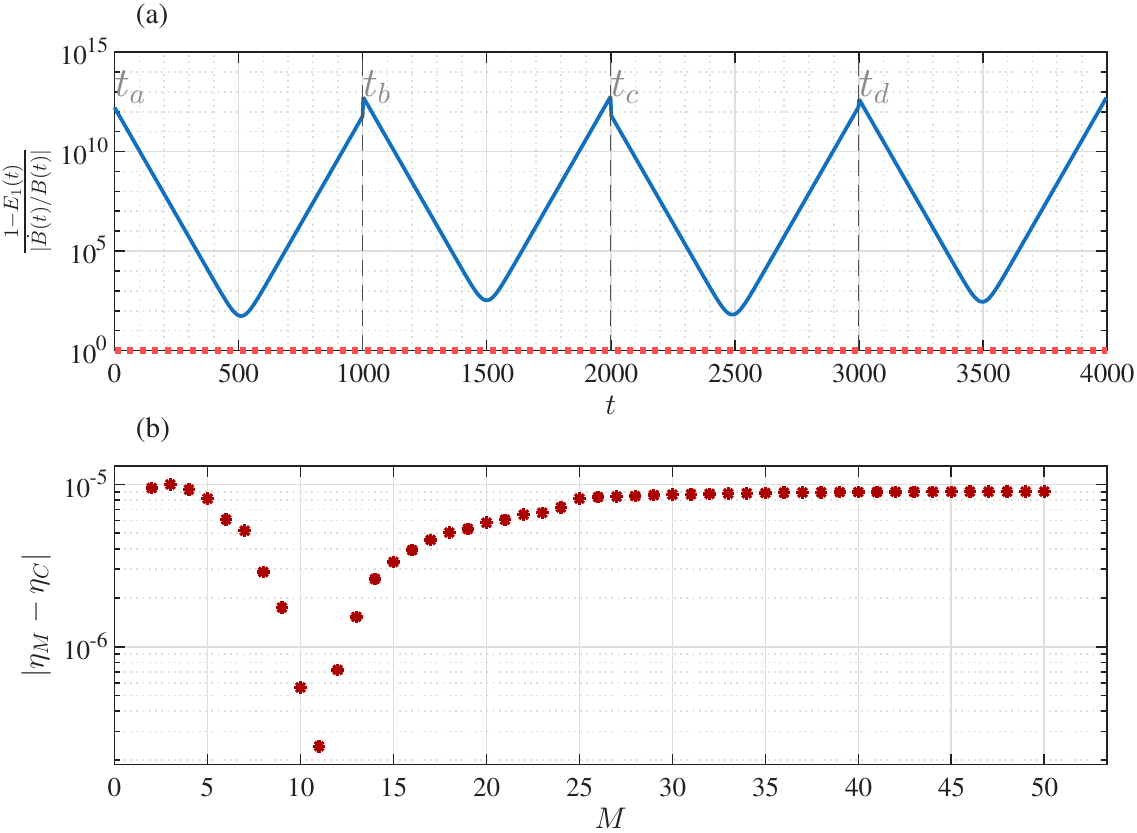}
    \caption{Adiabaticity and mesoscopic convergence of the soliton engine. (a) Adiabatic ratio $(1-E_1(t))B(t)/\vert\dot{B}(t)\vert$ (solid blue line) during the engine cycle. Vertical dashed lines separate the four strokes. The ratio remains well above unity (dashed red line), consistent with the adiabatic evolution of the spectral dynamics. (b) Absolute difference $\vert\eta_M-\eta_C \vert$ as a function of mode number,  between the mesoscopic and the classical efficiency.  \label{Fig:03}}
\end{figure}

The two-level reduction further requires adiabatic decoupling of the excited bound state from the continuum \cite{Kato1950, Kivshar1989, Avron1999, Kato2013}. Slow deformation of the soliton induces bound-continuum transitions whose amplitude is set by the relative deformation rate $\dot{B}(t)/B(t)$ and suppressed by the gap to the threshold $\Delta_1(t)=1-E_1(t)$. The instantaneous-mode expansion of the linearized fluctuation dynamics in our system yields the adiabaticity condition \cite{CompanionArticle}
\begin{equation}
    \label{Eq:06}
    1-E_1(t)\gg\frac{\vert\dot{B}(t)\vert}{B(t)}\,,
\end{equation}
which guarantees negligible radiative leakage into scattering states. Figure~\ref{Fig:03}(a) demonstrates that the condition of Eq.~\eqref{Eq:06} holds throughout the whole Carnot cycle in our numerical simulations. Violation of this condition causes the internal mode to hybridize with the continuum, leading to phonon emission and irreversibility. Full derivations are provided in the Companion Article \cite{CompanionArticle} and its appendices. 

Finally, we construct a mesoscopic Galerkin-based model of the efficiency by projecting the velocity field onto a finite set of $M$ dominant modes $\left\{\Phi_n\right\}_{n=0}^{M}$ obtained through proper orthogonal decomposition \cite{Berkooz1993}, $\partial_t\theta=v_M+v_{\perp}$. Here, $v_M$ belongs to the resolved subspace spanned by the first $M$ modes and $v_{\perp}$ represents unresolved degrees of freedom. This procedure is analogous to coarse-graining in the statistical physics of fields \cite{Kardar2007-2} and separates the resolved working medium from the unresolved bath degrees of freedom, thereby allowing the efficiency to be evaluated at intermediate levels of description. The mesoscopic kinetic energy of the working medium is $U_M(t):=(1/2)\int\mbox{d}x\, v_M^2$, and the projected energy balance yields $\mbox{d}U_M/\mbox{d}t=P_{\text{drive}}^{(M)}-P_{\text{diss}}^{(M)}-\Pi_{M\to\perp}$, where $\Pi_{M\to\perp}$ represents energy transferred to unresolved modes. The efficiency of the mesoscopic engine becomes
\begin{equation}
    \label{Eq:07}
    \eta_M=1-\frac{Q_{\text{diss}}^{(M)}+Q_{\text{leak}}^{(M)}}{Q_{\text{in}}^{(M)}}\,.
\end{equation}

The efficiencies obtained from Eqs.~\eqref{Eq:05} and~\eqref{Eq:07} are compared in Fig.~\ref{Fig:03}(b). The full nonlinear field description exhibits a high efficiency, reflecting the weak dissipation of the soliton dynamics under slow spectral deformation. More importantly, the comparison across different levels of description reveals a qualitative effect: few-level models fail to capture a significant fraction of the reversible energy storage present in the extended soliton structure. When only a small number of modes are retained, substantial energy leakage occurs into unresolved degrees of freedom. As more modes are included, the resolved system progressively captures the internal spectral dynamics of the soliton and the mesoscopic efficiency approaches the full-field result.

At first, it might seem that the spatial fluctuations of an extended object, like a soliton, will contribute to more dissipation channels to the system and thus decrease the efficiency of the engine. Our results, however, reveal that this expectation does not strictly hold. Figure~\ref{Fig:03}(b) shows that $\eta_M$ initially decreases as the number of modes increases from $M=2$ to $M=11$, in line with the naive expectation.  However, for $M>11$ the mesoscopic efficiency increases, and the mesoscopic model progressively approaches the result of the classical field description as more modes are resolved, with $\eta_M\to\eta_C\simeq0.9998$. In contrast, the spectral two-level model systematically underestimates the engine efficiency, $\eta_{\text{TLS}}\simeq0.24$. The discrepancy arises because the two-level model neglects several energy channels that are intrinsic to the soliton as an extended nonlinear structure. In full-field dynamics, energy can be stored not only in discrete bound states but also in deformation and radiation modes, as well as in nonlinear interactions among them. The mesoscopic model provides a quantitative interpretation of this effect. These findings demonstrate that extended coherent structures can outperform few-level working media, because their internal spectral structure provides additional pathways for storing and redistributing energy. Solitons, therefore, constitute a new class of thermodynamic working substances whose properties lie between microscopic quantum systems and macroscopic classical media. 

In conclusion, our results establish solitons as a new class of thermodynamic working substance in heat engines whose performance is intrinsically tied to their nonlinear and extended nature. Unlike conventional few-level engines, where energy exchange is restricted to a small set of discrete states, a soliton combines particle-like robustness with an internal spectral structure that couples discrete modes, continuum radiation, and nonlinear field dynamics. This richer structure enables additional pathways for storing and redistributing energy, leading to thermodynamic behavior that cannot be captured by simple few-level models. Beyond the specific implementation considered here, these findings suggest a broader principle: extended coherent excitations can enhance thermodynamic performance by exploiting internal spectral degrees of freedom. Our work, therefore, opens a route toward heat engines based on nonlinear coherent structures and provides a framework for exploring thermodynamics in systems that lie between microscopic quantum devices and macroscopic classical media. 

\emph{Acknowledgments --}
The authors thank Leonardo Gordillo for fruitful discussions. This work was funded by the Agencia Nacional de Investigación y Desarrollo (ANID---Chile) through the Postdoctoral FONDECYT grant 3240443 (M.A.) and the FONDECYT de Iniciación grant 11251397 (J.F.M.).

%




 


\end{document}